\documentclass[prd,eqsecnum,noshowpacs,nofootinbib,preprintnumbers]{revtex4}
\usepackage{amsmath,amssymb}
\usepackage[dvips]{graphicx,color}

\input epsf
\usepackage{color,graphics,latexsym,amsfonts}

\begin{document}
\title{Cosmological solutions of massive gravity on de Sitter}
\author{David  Langlois$^{1,2}$, Atsushi Naruko$^1$}
\affiliation{$^1$ APC (CNRS-Universit\'e Paris 7), 10 rue Alice Domon et L\'eonie Duquet, 75205 Paris Cedex 13, France;
\\
$^2$ IAP, 98bis Boulevard Arago, 75014 Paris, France
  }

\date{\today}

\def\beq{\begin{equation}}
\def\eeq{\end{equation}}
\newcommand{\bea}{\begin{eqnarray}}
\newcommand{\eea}{\end{eqnarray}}
\def\bi{\begin{itemize}}
\def\ei{\end{itemize}}
\def\ba{\begin{array}}
\def\ea{\end{array}}
\def\bfig{\begin{figure}}
\def\efig{\end{figure}}

\def\half{\mbox{\scriptsize{${{1}\over{2}}$}}}
\def\halff{\mbox{\scriptsize{${{5}\over{2}}$}}}
\def\quarter{\mbox{\scriptsize{${{1}\over{4}}$}}}
\def\eighth{\mbox{\scriptsize{${{1}\over{8}}$}}}

\def\mP{{m_P}}  
\def\bk{{\bf k}}
\def\bx{{\bf x}}
\def\a{\alpha}
\def\Hc{H_c}
\def\k{k}
\def\sf{\varepsilon_f}
\def\r{r}
\def\H{{\cal H}}

\newcommand\bmp[1]{\begin{minipage}[c]{#1\textwidth}}
\newcommand\emp{\end{minipage}}

\newcommand{\ma}[1]{{\mathrm{#1}}}
\newcommand{\calH}{{\cal H}}
\newcommand{\calL}{{\cal L}}
\newcommand{\calK}{{\cal K}}
\newcommand{\pa}{{\partial}}
\newcommand{\red}[1]{{\textcolor{red}{#1}}}

\begin{abstract}
In the framework of the recently proposed models of massive gravity, defined with respect to a de Sitter reference metric, we obtain new homogeneous and isotropic solutions for arbitrary cosmological matter and  arbitrary spatial curvature. These solutions can be classified into three branches. In the first two, the massive gravity terms behave like a cosmological constant. In the third branch, the massive gravity effects can be described by a time evolving effective fluid with rather remarkable features, including  the property to behave as  a cosmological constant  at late time.
\end{abstract}

\maketitle

\section{Introduction}
Long after the first attempt by Pauli and Fierz to give a mass to the graviton~\cite{Fierz:1939ix}, it has been realized, decades ago,
 that finding a healthy nonlinear massive extension of general relativity  represents a formidable  challenge because it requires to get rid of the so-called Bouldware-Deser  ghost~\cite{Boulware:1973my}.  Very recently, de Rham, Gabadadze and Tolley (dRGT) succeeded in constructing a massive theory of gravity that satisfies this criterion~\cite{deRham:2010kj}, as later confirmed in \cite{Hassan:2011hr}. Beyond its obvious theoretical interest, this achievement  has  a special significance in a context where most of the matter content of the Universe remains unknown and alternative explanations for dark energy and/or dark matter could reveal appealing. 
This explains why this recent model  has attracted a lot of  attention, especially for  its cosmological consequences. In this respect, a  surprising discovery was that  dRGT massive gravity does not allow for spatially {\it flat} homogeneous and isotropic solutions~\cite{D'Amico:2011jj}. However, {\it open} cosmological solutions were obtained, with two branches of solutions in which the massive graviton terms lead to an effective cosmological constant \cite{Gumrukcuoglu:2011ew} (other solutions relevant for cosmology can  be found in e.g. \cite{Koyama:2011xz,Koyama:2011yg,Gratia:2012wt,Kobayashi:2012fz,Volkov:2012cf}).

In the present work, we start from a slightly modified version  of the original dRGT massive gravity in which the (a priori arbitrary) reference geometry    is chosen to be de Sitter instead of Minkowski. A similar setting was explored very recently in \cite{deRham:2012kf} and \cite{Fasiello:2012rw}.
The de Sitter geometry possesses as many symmetries as the flat geometry but introduces  a mass scale $H_c$ as additional parameter.  In this setup,  we have been  able to find new  cosmological solutions with flat, open or closed spatial geometry, for  arbitrary cosmological matter.
 Our solutions can be classified in  three branches, two of which being analogous  to the open solutions of \cite{Gumrukcuoglu:2011ew}, while the last branch exhibits a new and rich  phenomenology.  

\section{Homogeneous and isotropic solutions of massive gravity}
We first present the theory of massive gravity introduced in \cite{deRham:2010kj}, which can be described in terms of the usual four-dimensional metric $g_{\mu\nu}$ and of four scalar fields $\phi^a$ ($a=0,\dots, 3$), 
called the St\"uckelberg fields.  Gravity is  governed by  the  action 
\begin{align}
 S_{\rm grav} &= M_{pl}^2 \int d^4 x \sqrt{- g} \left[ \frac{1}{2} R
 + m_g^2 \Bigl( \calL_2 + \alpha_3 \calL_3 + \alpha_4 \calL_4 \Bigr) \right] \,,
\end{align}
 where the first term is  the familiar Einstein-Hilbert Lagrangian 
 (we set $M_{pl}=1$ in the following)
 and the three additional terms are specific functions  of the metric  $g_{\mu\nu}$ and of the four scalar fields $\phi^a$, via  the tensor 
 \begin{gather}
 \label{K}
 \calK^\mu{}_\nu
 = \delta^\mu{}_\nu - \sqrt{f_{a b} \, g^{\mu\sigma}\pa_\sigma \phi^a \pa_\nu \phi^b}  \,,
 \end{gather}
 where $f_{ab}$ is called the reference, or fiducial, metric (the square root must be understood in a matricial sense). The explicit expressions for these additional terms in the Lagrangian are 
 \begin{align}
 \calL_2 &= \frac{1}{2} \Bigl( [\calK]^2 - [\calK^2] \Bigr) \\
 \calL_3 &= \frac{1}{6} \Bigl( [\calK]^3 - 3 [\calK] [\calK^2]
 + 2 [\calK^3] \Bigr) \\
 \calL_4 &= \frac{1}{24} \Bigl( [\calK]^4 - 6 [\calK]^2 [\calK^2]
 + 3 [\calK^2]^2 + 8 [\calK] [\calK^3] - 6 [\calK^4] \Bigr) \,  
 \end{align}
 where the standard matrix notation is used (i.e. $(\calK^2)^\mu_{\ \nu}=\calK^\mu_{\ \sigma}\calK^\sigma_{\ \nu}$) and the brackets  represent  a trace.

 We now restrict our discussion to  a  FLRW (Friedmann-Lema\^itre-Robertson-Walker) geometry, of arbitrary spatial curvature, described by the metric 
\begin{gather}
 ds^2 = g_{\mu \nu} dx^\mu dx^\nu
 = - N^2 (t) dt^2 + a^2 (t) \, \gamma_{i j} (x)  dx^i  dx^j,
\end{gather}
where the spatial metric $\gamma_{ij}$, written for example in terms of spherical coordinates, reads
\begin{gather}
 \gamma_{i j} (x) dx^i dx^j = \frac{dr^2}{1 - \k r^2} + r^2 \left(d \theta^2
 + \sin^2 \theta\, d \phi^2\right)
\end{gather}
with $\k=0$, $-1$ or $1$ for, respectively, flat, open or closed cosmologies.
 
In the present work, we take for the reference metric  $f_{ab}$ the de Sitter metric.  As we will see, and in contrast with the Minkowski case, one can easily construct flat, open and closed cosmologies by starting from the appropriate slicing of de Sitter.  Let us thus write the de Sitter metric in the form
\beq
f_{ab}\,d\phi^a \, d\phi^b=- dT^2+b^2_k(T) \, \gamma_{ij}(X) \, dX^i dX^j,
\eeq
where
 the functions $b_k(T)$ are defined by
\beq
\label{b_k}
b_0(T)=e^{\Hc T}, \qquad b_{-1}(T)=\Hc^{-1}\sinh(\Hc T), \qquad b_1(T)= \Hc^{-1}\cosh(\Hc T)\,.
\eeq
In the limit $\Hc\rightarrow 0$, one recovers the Minkowski metric in the flat and open cases: $b_0(T)=1$ and $b_{-1}(T)=T$,   the latter case corresponding to the  Milne metric for the flat geometry.

We must now specify the  St\"uckelberg fields so that the cosmological symmetries are satisfied. One sees immediately that the choice 
\beq
\phi^0=T=f(t), \qquad \phi^i=X^i=x^i
\eeq
leads to a homogeneous and isotropic tensor, 
\beq
f_{\mu\nu}=f_{a b}\,  \pa_\mu \phi^a\, \pa_\nu \phi^b={\rm Diag}\left\{-\dot f^2, b^2_k(f(t))\, \gamma_{ij}\right\}\,.
\eeq
Denoting $\sf$ the sign of $\dot f$, the corresponding matrix $\calK$, defined in (\ref{K}), is simply given by\footnote{We also assume $f>0$ in the case $k=-1$.}
\begin{align}
 \calK^0{}_0 = 1 - \sf\frac{\dot{f}}{N}  \,, ~~~~~
 \calK^i{}_j = \left( 1 - \frac{b_k(f)}{a} \right) \delta^i{}_j\,, ~~~~~
 \calK^i{}_0 = 0 \,, ~~~~~
 \calK^0{}_i = 0 \,.
\end{align}
Substituting in the Lagrangian of massive gravity, one gets
\begin{eqnarray}
{\cal L}_g&\equiv & \sqrt{-g}\left({\cal L}_2+\alpha_3  {\cal L}_3+\alpha_4 {\cal L}_4\right)
\cr
&=& \left(a- b_k(f)\right) \left\{N \left[a^2 (4 \alpha_3 +\alpha_4 +6)-a (5
   \alpha_3 +2 \alpha_4 +3) b_k(f)+(\alpha_3 +\alpha_4 ) b^2_k(f)\right]\right.
   \cr
  && \left. -\sf\,  \dot{f} \left[(3+3
   \alpha_3 +\alpha_4 )a^2- (3 \alpha_3 +2 \alpha_4 ) a\, b_k(f)+\alpha_4  b_k(f)^2\right]\right\}\,.
\end{eqnarray}
The equation of motion for $f(t)$ is obtained by varying this Lagrangian with respect to $f$:
\beq
 \left[ (3+3 \alpha_3 +\alpha_4) a^2-2  (1+2 \alpha_3 +\alpha_4 ) a \, b_k(f)+(\alpha_3 +\alpha_4 )
   b^2_k(f) \right] \left(\frac{\dot a}{N} - \sf\,  b'_k(f)\right)=0.
\eeq
In general, there are several solutions for $f$. The first two solutions correspond to 
\beq
\label{branch1}
b_k(f(t))= X_\pm \, a(t), \qquad X_\pm=\frac{1+2\a_3+\a_4\pm\sqrt{1+\a_3+\a_3^2-\a_4}}{\a_3+\a_4},
\eeq
which exist only if the function $b_k$ is invertible. For a Minkowski reference metric $f_{ab}=\eta_{ab}$, one sees immediatly that there is no solution in the flat case since $b_0(f)=1$, whereas $b_{-1}(f)=f$  leads to two branches of solutions  in the open case, in agreement with the conclusions of \cite{D'Amico:2011jj} and \cite{Gumrukcuoglu:2011ew}.

Let us now concentrate on the last  branch defined by the condition
\beq
\label{branch3}
\sf b_k'(f)= \frac{\dot a}{N}\,.
\eeq
It is non trivial  only if $b_k'$ is an invertible function,  which is not the case with a  Minkowski reference metric, either in the flat or open cases. However,  in our case,  one can obtain an explicit solution for $f(t)$ with the functions $b_k$ given in (\ref{b_k}). Before examining the flat case, let us stress that the solutions in this branch are necessarily accelerating as can be seen by taking the time derivative of (\ref{branch3}), which yields
\beq
\ddot a=  b_k''(f)\, |\dot f| >0 \qquad (N=1)\,.
\eeq
In the particular case $\k=0$,  on which we will focus in the following, one finds (assuming $\dot f>0$)
\beq
\label{f0}
f(t)=\Hc^{-1} \ln\left(\frac{H(t)\, a(t)}{\Hc}\right)\,,  \qquad  H\equiv \frac{\dot a}{Na}
\eeq
where $H$ denotes the usual Hubble parameter. 

\section{Friedmann equations and effective gravitational fluid}
To obtain the Friedmann equations, one must add to  ${\cal L}_g$ the usual Einstein-Hilbert term, which reads 
\beq
{\cal L}_{\rm EH}=-\frac{3\dot a^2 a}{N}+3\k N a\,,
\eeq
as well as an arbitrary  matter Lagrangian $\calL_m$ that describes ordinary cosmological matter. 
Variation of the total Lagrangian  with respect to the lapse function $N$ (which will be set to $1$ in the following)  then yields the first Friedmann equation
\beq
3H^2+3\frac{\k}{a^2}=\rho_m+\rho_g, \qquad H\equiv \frac{\dot a}{a},
\eeq
where $\rho_m$ denotes the ordinary  matter energy density whilst $\rho_g$ corresponds to an effective energy density arising from  the massive 
gravity action:
\beq
\label{rho_g_f}
\rho_g\equiv\frac{m_g^2}{a^3} \left(b_k(f)-a\right)\left\{(6+4 \alpha_3 +\alpha_4)\, a^2- (3+5 \alpha_3 +2\alpha_4 ) \, a \, b_k(f)+(\alpha_3 +\alpha_4 ) b^2_k(f) \right\}\,.
\eeq
The variation of the total action with respect to $a(t)$ yields the second Friedmann equation in the form
\beq
2\dot H+3H^2+\frac{\k}{a^2}=-P_m-P_g\,,
\eeq
 with the effective pressure 
\begin{eqnarray}
\label{P_g_f}
P_g&\equiv&\frac{m_g^2}{a^3} \left\{ \left(6+4 \alpha_3 +\alpha_4-(3+3\a_3+\a_4)\dot f\right) a^2-2  \left(3+3 \alpha_3 +\alpha_4-(1+2\a_3+\a_4)\dot f \right) a \, b_k(f) \right.
\cr
&&
\left. \qquad +\left(1+2\alpha_3 +\alpha_4-(\a_3+\a_4)\dot f \right) b^2_k(f) \right\}\,.
\end{eqnarray}
We now study the expressions of $\rho_g$ and $P_g$ for the three branches of solutions identified previously.

\subsubsection{First two branches}
Substituting the solution (\ref{branch1}), one finds that the massive gravity contribution behaves like a cosmological constant with
\beq
\rho_g=-P_g= -m_g^2 \frac{\left(1+\a_3\pm\sqrt{1+\a_3+\a_3^2-\a_4}\right)\left(1+\a_3^2-2\a_4\pm (1+\a_3)\sqrt{1+\a_3+\a_3^2-\a_4}\right)}{(\a_3+\a_4)^2}\,.
\eeq
Note that the terms proportional to $\dot f$ in (\ref{P_g_f}) cancel because they are proportional to the combination that appears in the equation of motion for $f$. We recover exactly the result of \cite{Gumrukcuoglu:2011ew}, even if the spatial curvature is no longer restricted to be negative. Remarkably, the result is independent of $\Hc$. 

\subsubsection{Third branch}
Let us now turn to the third branch where the effective gravitational fluid follows a much more sophisticated behaviour.  
For simplicity, we  consider here only the flat case, but it is straightforward to extend the following analysis to the open and closed cases.
Upon substituting the explicit solution (\ref{f0}) for $f$   into  (\ref{rho_g_f}) and (\ref{P_g_f}), one gets
\beq
\label{rho_g}
\rho_g=-m_g^2\left(1-\frac{H}{\Hc}\right) \left\{6+4 \a_3 +\a_4-(3+5 \a_3 +2 \a_4 ) \frac{H}{\Hc}
+(\a_3 +\a_4) \frac{H^2}{\Hc^2}\right\}
\eeq
and 
\begin{eqnarray}
P_g&=& m_g^2\left\{ 6+4\a_3+\a_4-(3+3\a_3+\a_4)\frac{H}{\Hc} \left(3+\frac{\dot H}{H^2}\right)+(1+2\a_3+\a_4)\frac{H^2}{\Hc^2} \left(3+2\frac{\dot H}{H^2}\right)
\right.
\cr
&& \left. \qquad \qquad 
-(\a_3+\a_4)\frac{H^3}{\Hc^3} \left(1+\frac{\dot H}{H^2}\right)\right\}\,.
\end{eqnarray}
It can be noticed that  (\ref{rho_g}) coincides with the expression  obtained by \cite{Fasiello:2012rw} in the special case of  de Sitter cosmology, i.e.  with a constant $H$, on a de Sitter reference  metric, although the Hubble parameter is time-dependent in our case.
One can check explicitly that  the effective gravitational fluid, characterized by $\rho_g$ and $P_g$,  satisfies  the usual conservation equation
\beq
\dot\rho_g+3H(\rho_g+P_g)=0\,.
\eeq
The behaviour of the effective fluid described by the above energy density and pressure is quite peculiar. The energy density $\rho_g$ can be positive or negative and its sign can change during time evolution, when the ratio $H/\Hc$ crosses specific values, which depend on the parameters $\a_3$ and $\a_4$. For example, in the minimal model with $\a_3=\a_4=0$, $\rho_g$ changes sign when $H=\Hc$ or $H=2\Hc$.

Combining the second  Friedmann equation with  the first one yields the relation
\beq
\left[m_g^2\frac{H}{\Hc}\left(3+3\a_3+\a_4-2(1+2\a_3+\a_4)\frac{H}{\Hc}+(\a_3+\a_4)\frac{H^2}{\Hc^2}\right)-2H^2 \right]\frac{\dot H}{H^2}=(1+w_m)\rho_m \,,
\eeq
where $w_m\equiv P_m/\rho_m$.
One can thus identify a critical value for $H$ when the term between brackets, of the form $\tilde m^2-2H^2\equiv m_g^2 \, \chi(H/\Hc)-2H^2$, vanishes. 
Remarkably, this critical value coincides with  the Higuchi bound derived in \cite{Fasiello:2012rw} for de Sitter cosmology. Moreover, one can see that for the  two other branches discussed earlier, the function $\chi(H/\Hc)$ vanishes, which seems to be related to the fact that the kinetic energy of the scalar mode around these solutions vanishes, as found in  \cite{Gumrukcuoglu:2011zh}, since the kinetic term is proportional to $\tilde m^2(\tilde m^2 -2H^2)$ according to \cite{Fasiello:2012rw}.  
If the Higuchi condition $\tilde m^2>2H^2$ is satisfied, the above relation implies that the Hubble parameter increases for matter satisfying the weak energy condition (i.e. $w_m>-1$ and $\rho_m>0$). 

Let us try to analyze the combined evolution of  the effective gravitational fluid with ordinary matter, in the minimal model where $\a_3=\a_4=0$ for simplicity. It is convenient to introduce the dimensionless quantities
\beq
Y\equiv \frac{H}{\Hc}, \qquad \r\equiv\frac{\rho_m}{\Hc^2}, \qquad \lambda\equiv \frac{m_g^2}{\Hc^2}\,.
\eeq
Using the first Friedmann equation to express $\rho_m$ as a function of $H$, one finds that the above equation can be rewritten as a differential equation for $Y$ and the full system can be written  in the form
\beq
\label{system1}
Y'=\frac{3(1+w_m) \left[(1+\lambda)Y^2-3\lambda Y+2\lambda\right]}{3\lambda-2(1+\lambda)Y}, \qquad r= 3 \left[(1+\lambda)Y^2-3\lambda Y+2\lambda\right]
\eeq
where a prime denotes a derivative with respect to the number of e-folds, i.e. $\dot Y=H Y'$. The second relation is simply a constraint between the value of the matter energy density and the Hubble parameter. In the following, we will only assume that the cosmological matter is characterized by $r>0$ and $w_m>-1$.
It is then worth noting that the Higuchi condition $\tilde m^2 -2H^2>0$ corresponds to
\beq
{\cal H}\equiv 3\lambda-2(1+\lambda)Y=-2(1+\lambda)(Y-Y_H)>0, \qquad Y_H\equiv \frac{3\lambda}{2(1+\lambda)}\,.
\eeq
In order to satisfy the Higuchi bound, one must therefore have $ Y<Y_H$ if $\lambda>-1$, or $Y> Y_H$ if $\lambda<-1$.

It is also useful to introduce the two roots of the numerator of the equation for $Y$,
\beq
Y_\pm=\frac{3\lambda\pm\sqrt{\lambda(\lambda-8)}}{2(1+\lambda)},
\eeq
which are defined if $\lambda>8$ or $\lambda<0$. 
Rewriting the dynamical system (\ref{system1}) in the form
\beq
Y'=-\frac32(1+w_m)\frac{(Y-Y_+)(Y-Y_-)}{Y-Y_H}, \qquad r= 3 (1+\lambda)(Y-Y_+) (Y-Y_-)>0\,,
\eeq
it is easy to study its evolution, depending on the value of $\lambda$:
\begin{itemize}
\item $\lambda >8$ (which implies $0< Y_-< Y_H<Y_+$): if $\H>0$, then $Y<Y_-$ and $Y$ tends towards $Y_-$ asymptotically. By contrast, if the Higuchi bound is not satisfied, i.e. $\H<0$, one must have $Y>Y_+$ and $Y$ decreases, converging asymptotically towards $Y_+$. 

\item $0<\lambda <8$ ($Y_+$ and $Y_-$ are not defined): the Higuchi bound is satisfied if $Y<Y_H$ initially, and $Y$ increases to reach $Y_H$ in a {\it finite} time. By contrast, if $Y>Y_H$ initially, the Higuchi bound is not satisfied and $Y$ decreases to reach $Y_H$ in a finite time. 

\item $-1<\lambda <0$ ($Y_H<0$): the Higuchi condition is never satisfied. The condition $r>0$ imposes $Y>Y_+$ and $Y$ decreases toward $Y_+$ asymptotically. 

\item $\lambda < -1$ (which implies $0< Y_+< Y_H<Y_-$): $\H>0$ imposes $Y_H<Y< Y_-$ initially and $Y$ tends asymptotically towards $Y_-$. If $\H<0$, one must have $ Y_+< Y< Y_H$ and $Y$ decreases towards $Y_+$. 

\end{itemize}
We thus find that in most  cases ($\lambda<0$ or $\lambda>8$), the effective gravitational energy density tends to a constant asymptotically, while the cosmological evolution approaches de Sitter, with a Hubble parameter that depends on $\lambda$ and is proportional to $\Hc$. When $0<\lambda<8$, the system evolves towards a singularity at  finite time. One can proceed similarly for general values $\a_3$ and $\a_4$ but the analysis is more complicated because the numerator and denominator of the equation for $Y$ become respectively third-order and second-order polynomials in $Y$.

\section{Conclusion}
In the present work, we have obtained spatially flat (as well as open or closed) FLRW  solutions with arbitrary cosmological matter in the context of ghost-free models of massive gravity, evading the no-go theorem of \cite{D'Amico:2011jj} by adopting  a de Sitter  reference metric instead of Minkowski.   The constraint equation for the St\"uckelberg fields leads to three branches. In  two branches, one finds that the effective gravitational fluid behaves like a cosmological constant, whose value, remarkably,   is independent of $\Hc$ and coincides exactly with the value obtained in \cite{Gumrukcuoglu:2011ew} for the specific case of open FLRW solutions with Minkowski as  reference metric. 
By contrast, the third branch  exhibits a much  richer phenomenology, although  expanding cosmological solutions are restricted to be accelerating. The  massive gravity effects can be described by  an effective fluid, which is in general time-dependent since its energy density depends on the physical Hubble parameter $H$ (and its pressure on $\dot H$ as well). In the simplest case where $\a_3=\a_4=0$ we have investigated the cosmological evolution and found that the outcome is either a singularity at finite time or a de Sitter evolution, depending  on the value of the ratio $\lambda=m_g^2/\Hc^2$. 

To conclude, massive gravity on de Sitter  leads to new  solutions with surprising features. It would  be worth exploring further these solutions, in particular by investigating more systematically the parameter space for the coefficients $\a_3$ and $\a_4$. It would also be interesting  to study perturbations around  these new solutions, by extending previous works on this topic (see e.g. \cite{Gumrukcuoglu:2011zh,Fasiello:2012rw}).

\acknowledgements{We would like to thank Shinji Mukohyama for  very instructive discussions, 
as well as Matteo Fiasello, Kazuya Koyama, Eric Linder and Andrew Tolley for  helpful remarks.
 A.N.  was supported by JSPS Postdoctoral Fellowships for Research Abroad. D.L. was partly supported by the ANR (Agence Nationale de la Recherche) grant ÒSTR-COSMOÓ ANR-09-BLAN-0157-01.}

\end{document}